\documentclass[useAMS,usenatbib]{mnras}
\usepackage{graphicx} 
\usepackage{subcaption}
\usepackage{amsmath} 
\usepackage{xcolor}
\usepackage{float}
\usepackage{caption}
\usepackage{placeins}
\usepackage{lipsum}
\usepackage{multicol}
\usepackage{soul}
\usepackage{amssymb}
\usepackage{ragged2e}
\usepackage{threeparttable}
\def\refjnl#1{{\rm#1}}
\def\aj{\refjnl{AJ}}                   
\def\apj{\refjnl{ApJ}}                 
\def\apjl{\refjnl{ApJ}}                
\def\apjs{\refjnl{ApJS}}               
\def\ao{\refjnl{Appl.~Opt.}}           
\def\aap{\refjnl{A\&A}}                
\def\mnras{\refjnl{MNRAS}}             
\def\pasp{\refjnl{PASP}}               
\def\pasj{\refjnl{PASJ}}               
\def\ssr{\refjnl{Space~Sci.~Rev.}}     

\def\be{\begin{equation}} 
\def\ee{\end{equation}}

\def\HeII{\hbox{He~$\scriptstyle\rm II\ $}}

\def\gsim{\lower.5ex\hbox{\gtsima}} 
\def\lsim{\lower.5ex\hbox{\ltsima}} \def\gtsima{$\; \buildrel > \over 
\sim \;$} \def\ltsima{$\; \buildrel < \over \sim \;$} \def\prosima{$\; 
\buildrel \propto \over \sim \;$} \def\gsim{\lower.5ex\hbox{\gtsima}} 
\def\lsim{\lower.5ex\hbox{\ltsima}} 
\def\simgt{\lower.5ex\hbox{\gtsima}} 
\def\simlt{\lower.5ex\hbox{\ltsima}} 
\def\simpr{\lower.5ex\hbox{\prosima}} \def\la{\lsim} \def\ga{\gsim} 
  
 \def\gtsima{$\; \buildrel > \over \sim \;$} 
\def\ltsima{$\; \buildrel < \over \sim \;$} 
\def\gsim{\lower.5ex\hbox{\gtsima}} 
\def\lsim{\lower.5ex\hbox{\ltsima}} 
\def\simgt{\lower.5ex\hbox{\gtsima}} 
\def\simlt{\lower.5ex\hbox{\ltsima}} 
\def\simpr{\lower.5ex\hbox{\prosima}} 
\def\la{\lsim} 
\def\ga{\gsim}

\def\E3{{\cal E}_{\rm g}^{III}}

\def\M*{M_*}
\def\Z*{Z_*}
\def\L*{L_*}

\def\ha{H$\alpha$}
\def\hb{H$\beta$}

\usepackage{soul}

\title[]{Markarian 590: The AGN Awakens}
\author[B.\ Palit et al.]{
Biswaraj Palit $^{1}$\thanks{bpalit@camk.edu.pl},
Marzena \'Sniegowska$^{2}$,
Alex Markowitz$^{1}$,
Agata R\'o\.za\'nska$^{1}$,
Joseph Farah$^{3,4}$
\newauthor
D. Andrew Howell$^{3,4}$
\\
$^{1}$Nicolaus Copernicus Astronomical Center, Polish Academy of Sciences, ul.\ Bartycka 18, 00-716 Warsaw, Poland\\
$^{2}$School of Physics and Astronomy, Tel Aviv University, Tel Aviv 69978, Israel\\
$^{3}$Las Cumbres Observatory, 6740 Cortona Drive, Suite 102, Goleta, CA 93117, USA\\
$^{4}$ Physics Department, University of California, Santa Barbara
Santa Barbara, CA, USA
}
\date{\today}
\pubyear{2021}
\begin{document}
\label{firstpage}
\pagerange{\pageref{firstpage}--\pageref{lastpage}}
\maketitle
\begin{abstract}
Changing-Look AGN (CLAGN) Mkn 590 recently underwent a sudden \lq re-ignition\rq, marked by substantial increases in optical/UV and X-ray continuum flux since last couple of years. \textit{Swift}-XRT observations revealed the re-emergence of a soft X-ray excess (SXE) as the source transitioned from a low-flux state in July 2023 to a significantly higher flux state in October 2024. This evolution was in response to an order-of-magnitude increase in extreme-UV (EUV) continuum emission, detected by \textit{Swift}-UVOT. Follow-up optical spectra from FLOYDS/Faulkes confirmed the enhancement of dynamically broadened Balmer lines, He II emission, and Fe II complex. As the Eddington fraction increased by a factor of $\sim$ 20 over the last 20 months, we found clear evidence of formation of a warm corona, strongly linked to the cold accretion disc underneath. Based on our multi-wavelength study  on recent data, we propose that Mkn 590 is currently becoming a Seyfert-1.2, similar to its state in 1990s. 
\end{abstract}
\section{Introduction}
\label{sec:introduction}
Changing-look AGN (CLAGN) exhibit dramatic changes in luminosity coupled with significant transformations in the Doppler-broadened optical Balmer emission lines \citep{ricci2023NatAs...7.1282R}. These phenomena likely arise from extreme shifts (by orders of magnitude) in the accretion rate onto the supermassive black hole (SMBH), and hence present our first observational clues into constraining the stochastic nature of AGN duty cycles \citep{mcleod2012ApJ...753..106M,Sch2015MNRAS.451.2517S}. Such variations in accretion could indicate modifications of the global supply of gas that may be associated with e.g., “chaotic cold accretion” clouds \citep{Gaspari2017MNRAS.466..677G}, or localized instabilities in the accretion disc which feeds the SMBH \citep[e.g.,][]{Noda2018MNRAS.480.3898N,marzena2020A&A...641A.167S}. The variations in accretion produce extreme variations in bolometric and ionizing luminosity illuminating the Broad Line Region (BLR), causing broad optical Balmer lines to appear/vanish \citep{LaMassa2015ApJ...800..144L,Ma2023ApJ...949...22M,Wu2023ApJ...958..146W}. CLAGNs’ luminosity variations provide a window into the geometry and mutual interactions of innermost accretion structures such as the accretion disc, the BLR, and the hot and warm X-ray coronae \citep[e.g.][]{gronkiewicz2023,Li2024ApJ}. For example, the BLR may form via dusty winds driven from the disc \citep{Czerny2011A&A...525L...8C,naddaf2022A&A...663A..77N}, so a change in disc luminosity or density may impact BLR structure in addition to its illumination.

CLAGN may also yield insights into the soft X-ray excess (SXE), a
nearly-ubiquitous component that emerges above the hard X-ray powerlaw (PL)
continuum below 2~keV.  Among several competing interpretations, Compton upscattering of disc photons by a warm,
optically-thick corona offers the most plausible scenario \citep{rozanska2015,POP2018A&A...611A..59P,zoghbi2023ApJ...957...69Z,ghosh2022ApJ...937...31G}. While it is still unclear as to the formation mechanism of such a warm
corona, it may empirically be present only at accretion rates relative to
Eddington ( $\lambda_{\rm Edd}$) above $\gtrsim 0.02$ 
  \citep{hagen2024MNRAS.534.2803H}. Variability in the SXE may be connected to that of the the
disc's optical/UV thermal emission \citep[e.g.,][] {mehdipor2011A&A...534A..39M}, the
presence of which may also depend on Eddington limited accretion rate ($\dot{m}_{\rm Edd}$) \citep{Nagar2005A&A...435..521N,hagen2024MNRAS.534.2803H}. These observations point to structural changes occurring
in the accretion disc near log($\lambda_{\rm Edd}$) $\sim$ $-$2, similar to the
behavior of black hole X-ray Binaries \citep[e.g. ][]{Wu2008ApJ...682..212W,Yang2015MNRAS.447.1692Y}. Sustained
monitoring of CLAGNs as they transition across this threshold thus may
yield insight into major structural changes in the disc \citep{Noda2018MNRAS.480.3898N,Lyu2021MNRAS.506.4188L}, as well as long-sought-after constraints on basic
energy requirements of warm corona formation.
\begin{figure*}
    \centering
    \includegraphics[height=6cm,width=18cm]{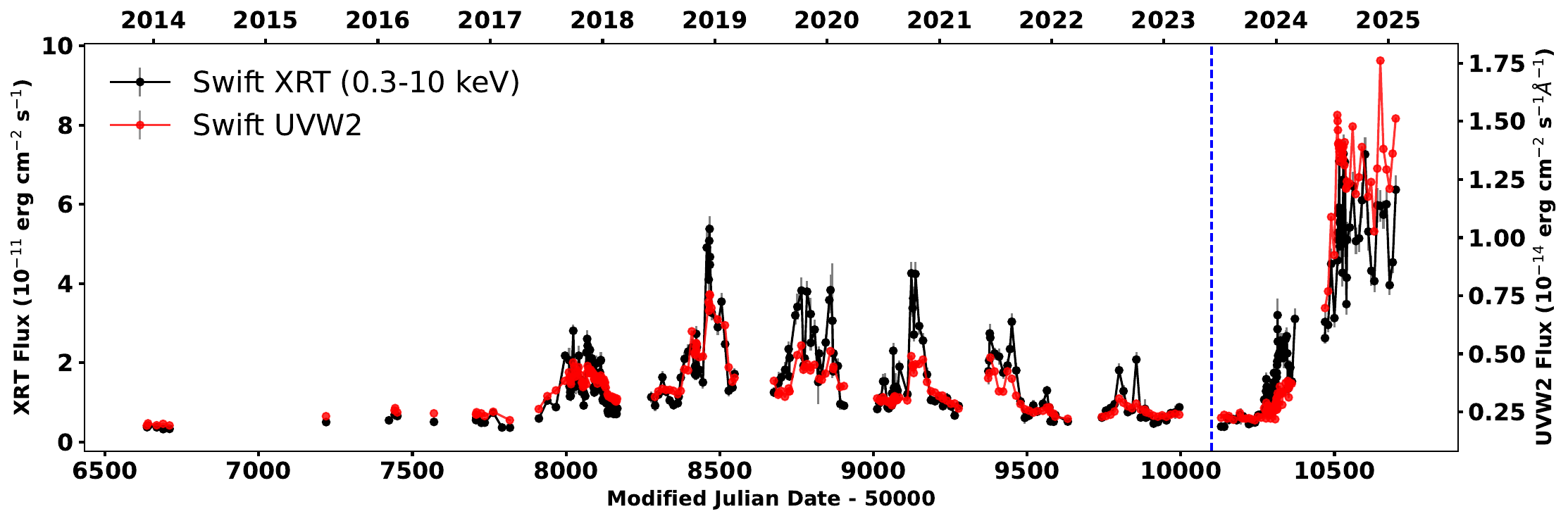}
    \caption{\textit{Swift}--XRT and {\bf UVW2} light curves between  2013 -- 2025. The {\bf UVW2} light curve (in red) is normalized with respect to mean XRT flux. Also, both Y axes share the same scale for direct comparability. The XRT absorption-corrected light curve between 0.3--10.0 keV is shown in black with 1\,$\sigma$ error bars in gray. The data beyond the blue dashed line constitutes the main focus for this paper. }
    \label{fig:xrt_full}
\end{figure*}
The nearby Seyfert Mkn 590 ($z$=0.0264) has displayed several major changing-look events over the course of 50 years \citep[and references therein]{Denney2014ApJ...796..134D}. In the 1970s, its spectrum was classified as sub-type 1.5 characterized by modest broad H$\beta$ line strength. However, its luminosity increased by a factor of 10 by the 1990s and spectra changed to sub-type 1 (very strong broad H$\beta$). From roughly 2000 through the mid 2010s, luminosity across all bands dropped by a factor of 100 and the source reverted back to a sub-type 1.9. \citet{mathur2018ApJ...866..123M} noted the presence of \ion{Mg}{ii} [$\lambda\lambda$2803,2796] during the low-flux state, so at least some component of the BLR remained apparently intact and illuminated. During 2017--2018, optical spectroscopy by \citet{Raimundo2017MNRAS.464.4227R}, using VLT/MUSE, and by \citet{mandal2021MNRAS.508.5296M}, using Subura/HDS), indicated the presence of broad Balmer lines, eventhough the continuum emission in the source was still weak. Starting in 2017--2018, Mkn~590 displayed temporary flaring behavior in both X-rays and the optical/UV for several years \citep{lawther2023MNRAS.519.3903L}. Interestingly, the SXE, which had completely vanished between 2004 and 2011, 
had still not yet re-appeared by 2020--2021 \citep[][using \textit{NuSTAR} and \textit{Swift}]{ghosh2022ApJ...937...31G}.

In this Letter, we report a significant rise in X-ray-UV-optical luminosity in Mkn 590 over the last year -- to unprecedented levels -- accompanied by a simultaneous, drastic increase in SXE flux and 
increases in the fluxes of broad Balmer lines, the \ion{Fe}{ii} complex, and \ion{He}{ii} emission.
Mkn 590 is thus an ideal laboratory to study how the various emission components in the innermost accretion flow are interconnected, and how they evolve as the source \lq\lq re-awakens\rq\rq. Here we adopt standard cosmological parameters i.e. Hubble constant $H_{\rm 0}$ = 69.6 km s$^{-1}$ Mpc$^{-1}$, mass density $\Omega_{\rm m}$ =0.3, and the density of dark energy as $\Omega_{\lambda}$=0.7.

 
\begin{figure}
   \centering
   \hspace{-.6cm}
    \includegraphics[height=7cm,width=9cm]{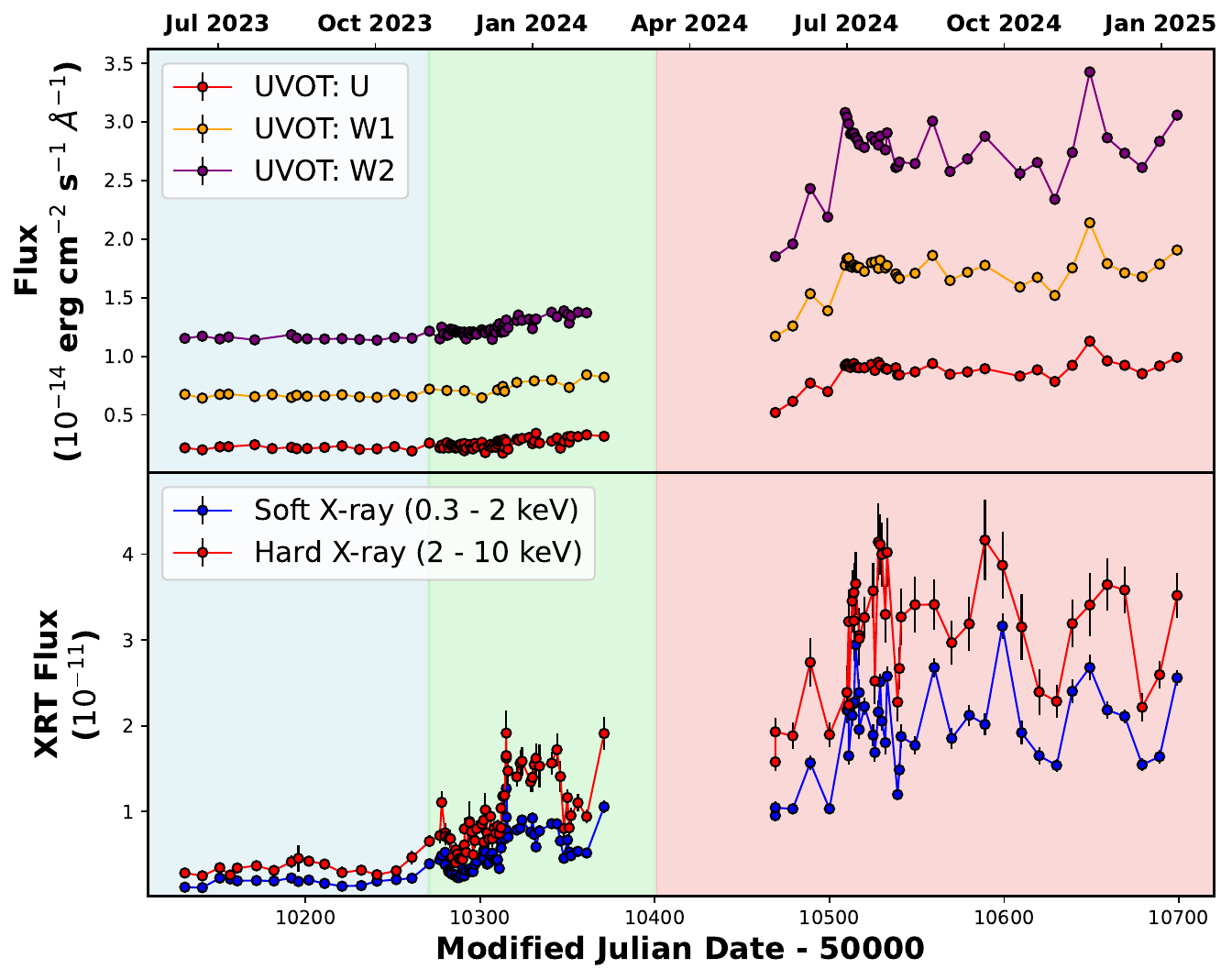}
    \caption{\textit{Swift}--UVOT (UVW2, UVW1 and U) lightcurves of Mkn 590 from Jun 2023 to Feb 2025, are shown in the upper panel. They are shifted by a constant flux factor for presentation purpose only. Bottom panel displays the \textit{Swift}--XRT light curve, separated into hard and soft X-ray energy range. X-ray fluxes are expressed in erg s$^{-1}$ cm$^{-2}$ and UVOT flux densities in erg s$^{-1}$ cm$^{-2}$ \AA$^{-1}$. The three lightly shaded regions ( blue,  green and red) correspond to our chosen epochs of interest for X-ray spectroscopy (see Tab.~\ref{tab:fits_fluxes}).}
    \label{fig:xrt}
\end{figure}
\section{\textit{Swift} Data}
\label{sec:sxs}
\begin{figure*}
    \centering
    \includegraphics[height=6.15cm,width=6.5cm]{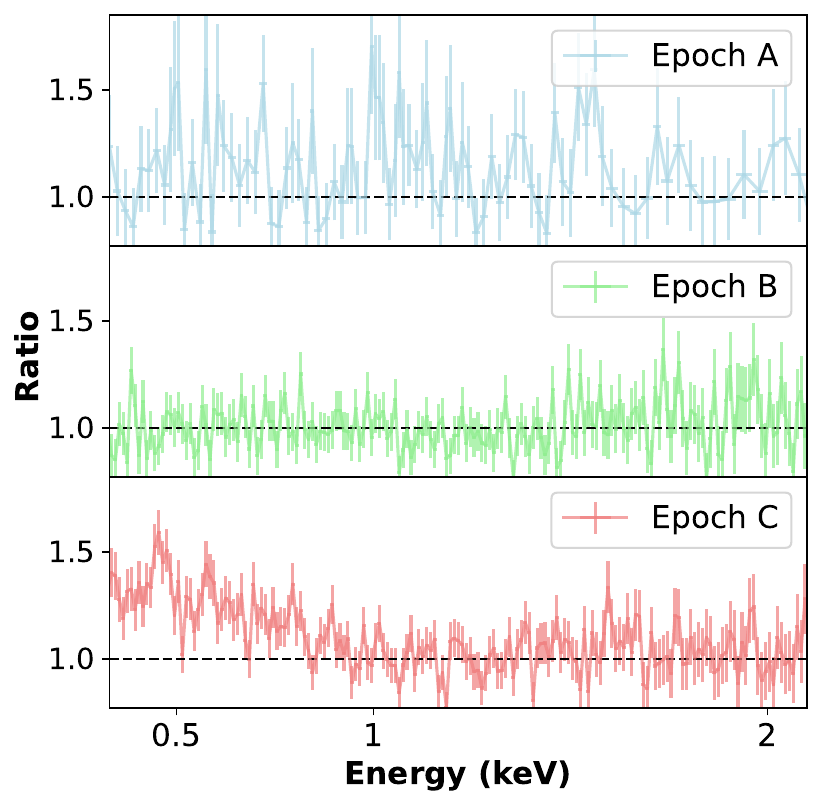}
\hspace{5mm}
\includegraphics[height=6.40cm,width=9.5cm]{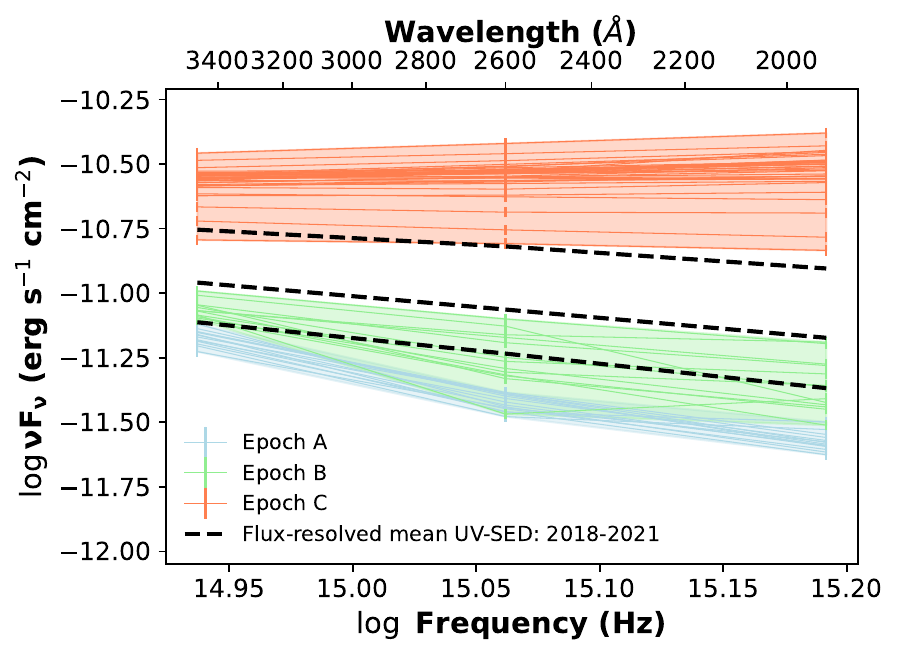}
    \caption{\textit{left:} Data/model residuals when a power law is fit only to the 2--10 keV band for epoch A, B, and C spectra, revealing the SXE's detection only in epoch C. 
    \textit{right:} UV SED, showing the change in slope from negative (at epochs A and B) to positive (epoch C) with increasing accretion rate. The bold dashed curves indicate averaged UV-SEDs between the period 2018-2021, divided into low, intermediate and high density bins ( mean flux densities roughly 2.4, 3.3 and 5.0 $\times$ 10$^{-15}$ erg s$^{-1}$ cm$^{-2}$ \AA$^{-1}$). In both panels, blue, green, and red denote epochs A, B, and C, respectively.}
    \label{fig:sxs}
\end{figure*}
In this paper, we consider data taken with the \textit{Neil Gehrels Swift Observatory} (\textit{Swift}) between 10 Dec.\ 2013 (starting ObsID: 37590002) and 02 Feb.\ 2025 (ending ObsID: 97768028) by both the X-ray Telescope \citep[XRT;][]{Burrows2005SSRv..120..165B} and the UV/Optical Telescope \citep[UVOT;][]{roming2005SSRv..120...95R}. In total, \textit{Swift} performed 365 XRT and UVOT (U, UVW1, and UVW2) observations simultaneously during this period, in particular with a high cadence ($\sim$ 1--2 days) since 2018; About 23 UVOT (U, W1 and W2) data corresponding to low sensitivity regions in the detector were neglected. 
The \textit{Swift}-XRT reduced data products were obtained from UK Swift Science Data Center (UK SSDC; \citet{Evans2009MNRAS.397.1177E}).
XRT spectra were extracted from a source region of 30$\arcsec$ with a background annulus spanning 40$\arcsec$--80$\arcsec$. 
UVOT photometry for all available filters were processed with \textsc{HEASoft} v6.32. 
Aperture photometry was performed with \textsc{uvotsource}, using a 5$\arcsec$-radius circle for the source and a background annulus spanning 35--75$\arcsec$; flux densities were corrected for Galactic reddening, assuming a reddening of $R$=3.1 \citep{cardelli1989ApJ...345..245C} and $E(B-V)$ = 0.0356 \citep{schpley2011ApJ...737..103S}.
The entire XRT light curve is displayed in Fig. \ref{fig:xrt_full} in black. We overlay-ed the UVW2 light curve (in red) after normalizing with respect to mean XRT flux. This letter mainly focuses on the \textit{Swift} data between Jun. 2023 to Feb. 2025, to the right of the blue dashed line, which has shown a  strong increase in UV and X-ray activity compared to previous years. These data are magnified in Fig.\ref{fig:xrt}.

Mkn 590 displayed a peak X-ray flux of $\sim$ 7 $\times$ 10$^{-11}$ erg s$^{-1}$ cm$^{-2}$ around August 2024, roughly 10 times the flux in July 2023. That July 2023 flux is comparable to the lowest X-ray flux ever detected in this source, which was during the historic low state of 2013 \citep{mathur2018ApJ...866..123M}. Among all the archival X-ray observations \citep[see papers:][]{Denney2014ApJ...796..134D,ghosh2022ApJ...937...31G,lawther2023MNRAS.519.3903L}, this August 2024 flux is the highest ever reported for this source; being comparable only to RXTE observations from $\sim$ 25 years ago, when the source displayed fluxes around 1-5 $\times$ 10$^{-11}$ erg s$^{-1}$ cm$^{-2}$ \citep{Alex2004ApJ...617..939M}. Furthermore, comparing the peak flux densities of U, UVW1 and UVW2 bands in our work with that of 2013 \citep{lawther2023MNRAS.519.3903L}, we note increases by factor of $\sim$ 7, 17 and 5, respectively. Meanwhile, comparing our peak flux densities with those in July of 2023, the increases were $\sim$ 4, 7 and 12 respectively.  Through the most recent observations, from Feb. 2025 (Fig. \ref{fig:xrt_full}, the source shows no sign of sustained decay). Overall, we captured the source undergoing a substantial and sudden rise in X-ray/UV flux over the last 1.5 years, a likely indication that the accretion rate has significantly increased. 

\begin{table}
    \centering
         
    \caption{\textit{Top:} Best-fitting X-ray spectral parameters for the absorbed PL model. Absorbed X-ray fluxes are in erg s$^{-1}$ cm$^{-2}$. $F_{\rm DBB}^{\rm abs}$ denotes the SXE flux from the absorbed \texttt{DBB+PL} model (see text for details). \textit{Bottom:} Corresponding UVOT flux densities. 
    }
    \begin{tabular}{|ccccc|}
    \hline
         PL model & Epoch A & Epoch B & Epoch C  \\
         \hline
         $\Gamma_{\rm hc}$ &1.62 $\pm$ 0.09 &1.69 $\pm$ 0.04 &1.69 $\pm$ 0.03 \\
         Norm ($\times$ 10$^3$) &0.66 $\pm$ 0.08 &2.08 $\pm$ 0.10 &6.63 $\pm$ 0.28 \\
         $\chi^{2}$/dof &55.25/44 &213.14/239 &249.67/264 \\ 
         \hline
         $F_{\rm 2 - 10}^{\rm abs}$ ($\times$ 10$^{-12}$) &3.25 $\pm$ 0.16 &8.58 $\pm$ 0.12 &27.07 $\pm$ 0.05 \\     
         \hline
         $F_{\rm DBB}^{\rm abs}$ ($\times$ 10$^{-13}$) &$<$ 0.37 &$<$ 0.28 &30.20 $\pm$ 0.06 \\
         \hline
         \hline
         \multicolumn{5}{|c|}{Mean UVOT flux densities ($\times$ 10$^{-15}$ erg s$^{-1}$ cm$^{-2}$ \AA$^{-1}$)} \\
         \hline
         U (3465 \AA) &2.19 $\pm$ 0.08 &2.53 $\pm$ 0.06 &8.85 $\pm$ 0.19 \\
         UVW1 (2600 \AA) &1.63 $\pm$ 0.08 &2.46 $\pm$ 0.07 &12.21 $\pm$ 0.23 \\
         UVW2 (1928 \AA) &1.44 $\pm$ 0.06 &2.40 $\pm$ 0.08 &17.50 $\pm$ 0.25 \\
         \hline
    \end{tabular}
    
    \label{tab:fits_fluxes}
\end{table}

To perform a time-resolved spectral analysis, we divided the 2023--2024 observing season into three sub-epochs (MJD 60131--60262, 60271--60372, and 60469--60708) and stacked their XRT-spectra. Each segment, referred to as epochs A, B and C, consisted of 16, 62, and 44 observations with stacked exposures of $\sim$ 35.8, 119.3 and, 64.4~ks, respectively. These epochs are marked as the blue, green and red patches, respectively, in Fig.~\ref{fig:xrt}, and denote low, intermediate, and flaring flux states. We fitted each spectrum over 2--10 keV with a power law (PL), absorbed only by the Galactic column of $N_{\rm H}$=2.77 $\times$ 10$^{20}$ cm$^{-2}$ \citep{Willingale2013MNRAS.431..394W}. 
Best-fitting parameters are listed in Tab.~\ref{tab:fits_fluxes}. 
The bottom panel of  Tab.~\ref{tab:fits_fluxes} contains mean flux densities for all available UVOT filters during each epoch. 
The hard X-ray spectral index $\Gamma_{\rm hc}$ does not vary significantly, while the 2--10~keV flux increased by a factor of $\sim$8 from epoch A to C. On extrapolating the PL to energies $\le$ 2~keV, we observe a rise in data/model residuals from epochs A/B to C -- the emergence of the SXE -- as displayed in Fig.~\ref{fig:sxs}. 
We fit a phenomenological disc-blackbody (DBB) + PL model in the 0.3--10 keV energy range to each spectrum, and we used the Akaike information criterion (AIC) \citep{liddle2007MNRAS.377L..74L}
to assess if this model is an improvement compared to a PL. The values of AIC increased for each of epochs A and B, but decreased drastically only for epoch C, with the fit improving at $\gg$ 99$\%$ confidence level. We list the 0.3--2.0 keV flux of the DBB component from epoch C, and upper limits to fluxes for epochs A/B, in Tab.~\ref{tab:fits_fluxes}.
The SXE flux in epoch C has increased by factors of $\geq$82 and $\geq$108
with respect to epochs A and B, respectively.
The upper limits in epochs A/B are far below the flux values measured by  \citet{ghosh2022ApJ...937...31G} in their detections of the SXE in 2002 and 2004, as well as lower than the upper limits\footnote{90$\%$ confidence limits both in this paper and in \citet{ghosh2022ApJ...937...31G}} during 2011--2021.
Meanwhile, the SXE flux in epoch C is greater than that in the 2002 and 2004 observations by factors of at least 7 and 8.1, respectively.
The ratio of 0.3--2.0 keV SXE flux to 2--10 keV PL flux during epoch C, $0.111\pm0.001$, is higher than in epochs A/B or any observation in \citet{ghosh2022ApJ...937...31G}.
For a more physically-motivated spectral fit, we replaced the DBB in epoch C with a Comptonized component modeled with \texttt{COMPTT}. We kept the seed photon temperature fixed at 2~eV and the optical depth fixed to 10 to mitigate parameter degeneracies. Our best fit yields an electron temperature of 0.44 $\pm$ 0.04 keV, similar to warm Comptonization modeling for numerous other Seyferts \citep{palit2024A&A...690A.308P}. It must be noted that the SXE feature could not be an artifact of ionized line of sight absorption mainly because of spectral softening and increase in soft X-ray flux at epoch C. Additionally, we tested Epochs A and B for full- or partial-covering obscuration by neutral or lowly-ionized gas in the line of sight, but column densities and/or covering fractions went towards zero, ruling out the possibility that the soft excess was present but obscured in Epochs A/B.

We constructed the UV spectral energy distributions (SEDs) for each epoch using U, UVW1 and UVW2 flux densities, plotted in the right panel of Fig.~\ref{fig:sxs}. Then, we divided the UVOT data between 2018-2021 into three flux density bins and in a similar way constructed their averaged flux-resolved UV-SEDs. The three SEDs are denoted by black dashed lines. The binning was based on the criteria that high flux state (top dashed lines) comprised the top 90\% of flux density values, low flux state (bottom dashed line) comprised the bottom 10\% and central 80\% to the intermediate flux state. Going from epoch A to epoch C, the 
UVW2 band has responded most strongly, rising by an order of magnitude since July 2023.Even compared to earlier epochs, the SED is noticeably flatter in epoch C, and its slope turns from negative to positive. As a caveat, though, we have not subtracted host galaxy contamination.
\begin{figure}
    \centering
    \includegraphics[scale=.38]{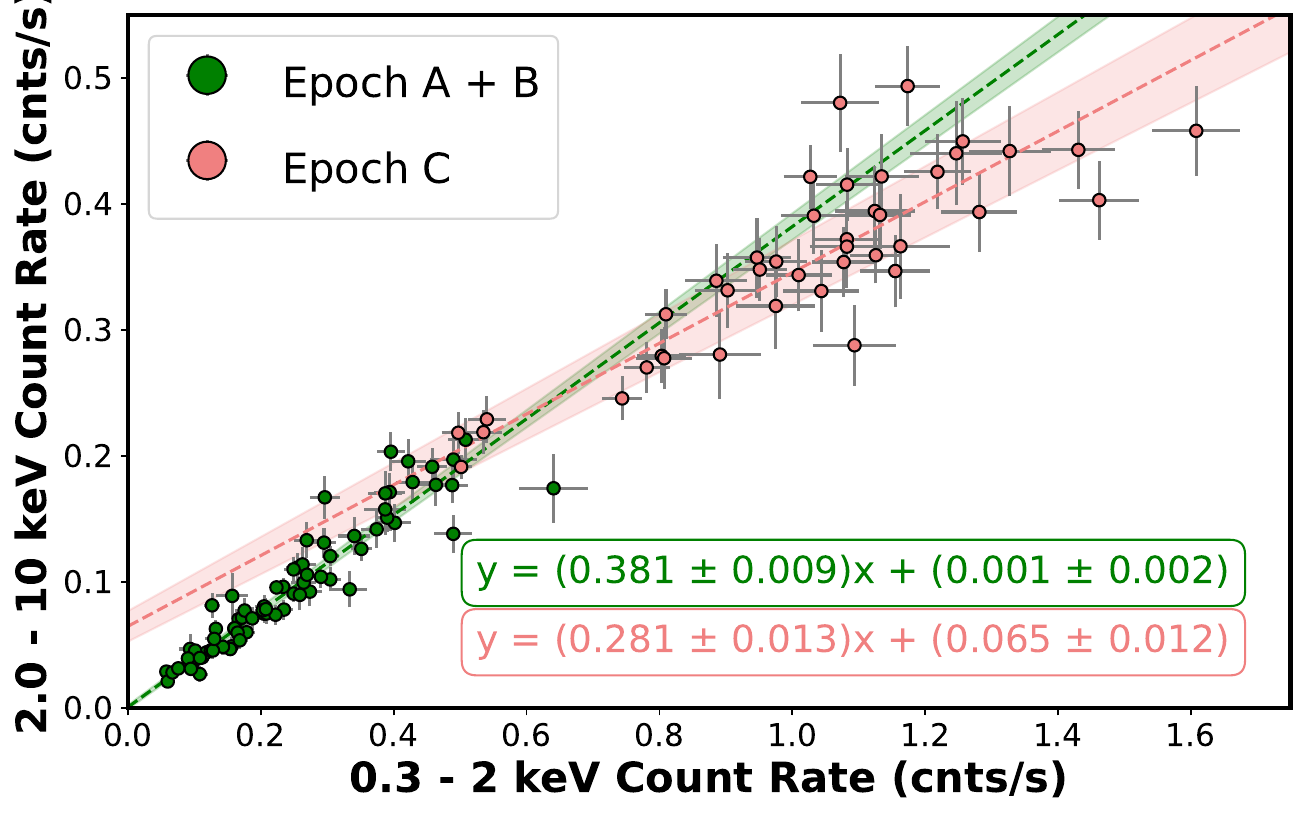}
    \caption{Count-Count plots for epochs of interest. The shaded areas represents 1\,$\sigma$ confidence interval.}
    \label{fig:cc}
\end{figure}

The Count-Count plot with Positive Offset (C3PO) technique offers a non-parametric approach to separate independently-variable continuum components \citep{Noda2011PASJ...63S.925N}. In Fig.~\ref{fig:cc}, we plot 2--10 keV versus 0.3--2.0 keV for each observation. We also plot the best-fit linear regression for epochs A/B and C, assuming an effective variance $\sigma_{\rm eff} = \sqrt{\sigma_{\rm y}^2 + a\,\sigma_{\rm x}^2}$. Here, $a$ denotes the slope of the best fit line, while $\sigma_{\rm x}$ and   $\sigma_{\rm y}$ are uncertainties in count rates. We note a clear divergence between epochs A/B and C, with epoch C showing a clear positive offset and flatter slope, indicating a disparate, independent source of emission of soft X-rays from that of hard X-rays. 
In summary, the SXE has come roaring back by epoch C, and is detected at its highest flux levels in over two decades.



\section{LCO Faulkes FLOYDS spectrum}
\label{sec:lco-floyds}

To assess the current BLR activity and emission characteristics, we obtained a 
new optical/near-IR spectrum using
the FLOYDS spectrograph \citep{Sand2011} at the Faulkes Telescope 
South, Siding Spring Observatory, through the Las Cumbres Observatory Globel Network (LCO).
The source was observed
on 8 Nov. 2024 (MJD 60622) with a  2$\arcsec$ slitwidth and a 1200~s exposure time.

\begin{figure}
    \centering
    \includegraphics[width=7.5cm,height=5.3cm]{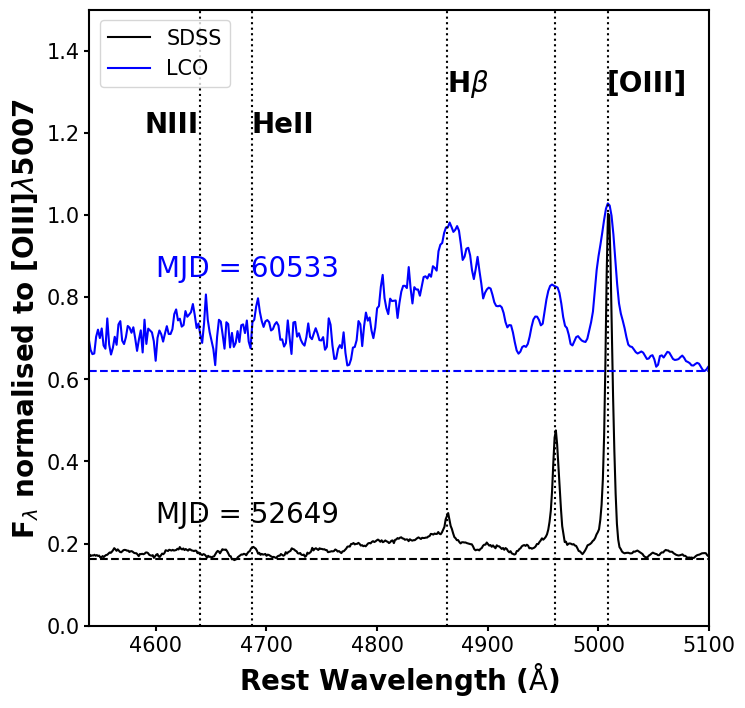}
    \caption{Comparison of optical spectra from LCO with that of SDSS, taken in 2003. The dashed lines denote monochromatic luminosity at 5100\AA.}
    \label{fig:w2}
\end{figure}

In Fig.~\ref{fig:w2}, we present a comparison of the H$\beta$ regions of the LCO Faulkes spectrum with an archival SDSS spectrum from 2003, when the source was undergoing a dimming state. 
The LCO Faulkes spectrum was greyscaled so that its integrated [\ion{O}{iii}]\,$\lambda5007$ flux matched that from the SDSS spectrum, following the framework of \cite{groningen1992}.
Of note are the asymmetric, blueward-skewed broad H$\beta$ line, increased broad H$\beta$ flux relative to [\ion{O}{iii}] flux in the 2024 spectrum, and
the stronger \ion{He}{ii} $\lambda$4687 emission in 2024. 
  %
 %
We modeled both spectra using {\sc PyQSOFit} \citep{pyqsofit}, assuming a PL\, emission component from an accretion disc, using an \ion{Fe}{ii} emission template from \citet{1992borosongreen}, and host galaxy template \citep{YIP2004AJ....128.2603Y}. We fit the whole wavelength range simultaneously. 
We used single Gaussians to model narrow lines of H$\alpha$, H$\beta$, [\ion{O}{iii}]~$\lambda\lambda$5007,4959 [\ion{N}{ii}]~$\lambda\lambda$6584,6548, [\ion{S}{ii}]~$\lambda\lambda$6731,6716, and \ion{O}{i}]$\lambda$6300. The \textcolor{black}{[\ion{O}{iii}]~$\lambda\lambda$5007,4959 doublet is modeled with single component.} 
For each broad Balmer line in both spectra, we used two broad Gaussians: one centered near rest-frame emission (hereafter BC), and one centered roughly 60--120 \AA\, lower, to model each blue wing (hereafter BL); all energy centroids were left free.
Finally, the \HeII\ line was modeled using a single BC.



Modeling of the spectrum is shown in Appendix~\ref{App:A}: Fig.~\ref{fig:SF} and the best-fitting parameters are listed in Table~\ref{tab:opt-spec-measurements}. Of note is that most components' fluxes in late-2024 are significantly higher than during 2003: the BC components of broad \ha\ and \hb\ increased in flux by factors of 9 and 6, respectively, while the BL components increased by factors of 1.2 for \ha\ and 4 for \hb.
\ion{He}{ii} $\lambda$4687 emission has increased by factor or 10. In addition,  the \ion{Fe}{ii} complex peaking near 4570 \AA\ is stronger in 2024. Both lines were routinely detected during the 1990s, but both were visibly absent between 2003 and 2014 \citep{Denney2014ApJ...796..134D}. 
{\textcolor{black}{However, there seems to be no major change in kinematic parameters between the two spectra:
both Balmer lines require two kinematic components in each spectrum,
and line widths and energy centroids have not evolved significantly.
}}
To classify Mkn~590's activity subtype, we used the ratio of broad H$\beta$ flux (sum of BC+BL) to  that [O\,{\sc iii}]\,$\lambda5007$, following \citet{Winkler1992MNRAS.257..659W}; we estimate the current subtype to be Seyfert 1.2, compared to being a subtype 1.5 in 2003.





\section{Discussion}
\label{sec:diss}

Mkn 590's recent extreme and sudden re-brightening in the UV and X-ray
bands, as tracked by \textit{Swift}, offers a rare opportunity to
observe an AGN re-illuminating its circumnuclear environment in real
time.
In this paper, we report a massive increase in the flux of the SXE component 
occurring sometime in the early 2024 timeframe, concurrent to strong increases
in UV (especially far-UV) flux. We also provide a new optical spectrum
of Mkn~590 as a reference for near-future spectroscopic studies; it
reveals a strengthening in flux of broad Balmer lines, as well as
\ion{He}{ii}, and \ion{Fe}{ii}.

The plateau-like nature of the light curves since mid-2024
in both X-rays and UV argues against the recent increases
being a Tidal Disruption Event \citep{vanvelzen2021ApJ...908....4V} or a Quasi Periodic Eruption \citep{guolo2024NatAs...8..347G}. Also, the absence of a strong [N III] $\lambda\,4640$ in the Faulkes spectrum was inconsistent with occurrence of Bowen Florescence Flares \citep{benny2019NatAs...3..242T,makry2023ApJ...953...32M}. Hence, we discuss our findings in the context of the \lq\lq re-awakening\rq\rq of the
accretion disk and warm and hot X-ray coronae.
Assuming an X-ray bolometric correction factor of 10  from
\citet{Duras2020A&A...636A..73D}, a luminosity distance of 115 Mpc
\citep{NED2006PASP..118.1711W}, and a SMBH mass of 4.75 $\times$
10$^{7}$ M$_{\odot}$ \citep{peterson1993ApJ...402..469P}, we estimate
$\lambda_{\rm Edd}$ = 0.01, 0.03, and 0.24 for Epochs A, B, and C,
respectively.
  The Faulkes spectrum yields a value of $\lambda_{\rm Edd}$ similar to
that of epoch C: $\lambda$log$L_{5100}$ is $6.03\times10^{43}$
erg~s$^{-1}$; assuming a bolometric correction factor of 9 from
\citet{ruonne2012MNRAS.422..478R}, we obtain $\lambda_{\rm Edd}$=
0.09 in Nov.\ 2024.

The flaring in the UV band and the emergence of the SXE is likely not 
coincidental.  In the context of warm Comptonization models, a
connection is expected
\citep{mehdipor2011A&A...534A..39M,POP2018A&A...611A..59P,parington2024arXiv241021432P}. One possible scenario is that the warm corona and SXE already existed
in all epochs, but in Epochs A and B, the warm corona was either
relatively starved of seed photons from the disk, or was simply
inefficient at producing substantial soft X-ray emission, and the soft
excess was simply drowned out by the $>$2 ~keV emission from the hot
corona.  Then, from Epochs A to C, as far-UV flux increased by a
factor of $\gtrsim$10, the SXE flux increased by a factor of $\geq$80
(with a non-linear response to far-UV flux), and it emerged above the hot
coronal component.

However, there seems to be a clear divergence in spectral variability
between epochs A/B and C, with the SXE's sudden increase between epochs B and C; in particular, we note that the ratio of 0.3--2.0 keV SXE flux to 2--10 keV PL flux jumped from $<0.003$ to 0.11. We thus
consider that there could have been
structural changes to the warm corona and/or the accretion disk.
It is generally thought that above and below a critical value of
$\lambda_{\rm Edd}$, typically log($\lambda_{\rm Edd}$) $\sim -2$, the
inner accretion flow in both Seyferts and in Black Hole X-ray Binaries
may transition between a geometrically-thin disk
and a radiatively-inefficient flow \citep{Esin1997ApJ...489..865E,Yang2015MNRAS.447.1692Y,Lyu2021MNRAS.506.4188L}.
In the context of the AGNSED model \citep{kubota2018MNRAS.480.1247K}, wherein the
warm corona exists in an annular region radially exterior to the hot
corona and interior to a standard geometrically-thin disk, an increase
in accretion rate leads to the enhancement of far-UV emission and the
appearance of a SXE around values of log(${\lambda}_{\rm Edd}$) $\sim$ $-$2 to $-$1.5
\citep{mitchell2023MNRAS.524.1796M,hagen2024MNRAS.534.2803H},
  not far from the values of ${\lambda}_{\rm Edd}$ inferred for Mkn~590 in this study.
In Mkn~590, the relatively stronger increase in UVW2 emission from epochs A to C is
consistent with thermalized emission from the disk annulus under the
warm corona becoming energetically more important than the emission
from the standard outer disc, while the warm corona expands radially,
increasing its soft X-ray output \citep{hagen2024MNRAS.534.2803H}.



It is also possible that the warm corona was weak in Epochs A/B, but
switched to a more dissipative state by Epoch C, characterized by a
large amount of heating at the photosphere of an inflated accretion
disc, which is now optically thick \citep{palit2024A&A...690A.308P}. Based on analytical studies by \citet{rozanska2015} and \citet{gronkiewicz2023}, rising magnetic dynamo in magnetically dominated disks could be responsible for such heating.
Furthermore. the radiation due to internal dissipation inside the warm
corona can be partly reprocessed by the cold passive disc underneath,
and re-emitted locally into the warm corona again
\citep{POP2020A&A...634A..85P}. This might have led to increased
production of UV seed photons for Compton up-scattering during epoch
C, leading to an increase in the ionizing luminosity impacting the
BLR.  Meanwhile, towards low accretion rates, an optically thick,
dissipative corona can collapse, leading to escape of seed photons, a
weak SXE, and dimming of EUV radiation. The exact mechanism of how
warm corona formation requires detailed broadband SED modeling for all
available archival data as well as new data, and constitutes a major
focus of our upcoming work.

Meanwhile, kinematic parameters for the broad Balmer emission lines are
comparable to those from optical spectra
from the mid-1990s \citep{peterson1998ApJ...501...82P}, 2003 and 2017 \citep{Raimundo2017MNRAS.464.4227R}.  
These features support the notion that a recent increase in accretion in the
disk has produced an enhancement of ionizing photons, illuminating and
exciting various regions of the BLR.  One scenario is that the BLR gas
has remained structurally stable over the last decades; currently, its
low-, intermediate- and high-ionization zones are being illuminated,
as evidenced by the enhanced \ion{Fe}{ii}, Balmer, and \ion{He}{ii} emission, respectively.
Similar responses of broad line fluxes to variations in luminosity
have been reported for numerous CLAGNs, e.g., NGC 1566
\citep{alloin1986ApJ...308...23A,Oknyn2019MNRAS.483..558O} and NGC
4151 \citep{shapalova2010A&A...509A.106S}.
However, we cannot rule out that some structural changes
may have occurred over the last couple decades.
The dynamical timescale at the radial location of the H$\beta$ BLR, 20
lt-dy \citep{peterson1998ApJ...501...82P}, is roughly 20 years.  Furthermore,
\citet{kubo2020MNRAS.491.4615K} note that a re-brightening event can
sublimate the innermost dusty regions in Mkn 590 within 4 light-years,
launching outflows that are observed in the mid-infrared bands and
that can contribute to optical broad line flux. 

The timescale for the SXE to increase massively in flux -- of order half a year between
epochs B and C -- is generally consistent with the thermal timescales
\citep[e.g.,][]{treves1988PASP..100..427T} expected for a standard thin disk at radii
of order 150 $R_{\rm g}$, and assuming a viscosity parameter $\alpha$
of order 0.1.
The increase in optical/UV luminosity may, speculatively,
also be associated with a propagating heating front
that increases $H/R$ and thus the local accretion rate
\citep{Ross2018MNRAS.480.4468R}.
Furthermore, replenishment of magnetic fields can be associated with
an approaching mechanical front which reactivates the inner accretion
disc by heating upper layers of the accretion disc, making the warm corona dissipative in nature.



Through this Letter, we report the re-invigoration of the CLAGN Mkn
590, which is showcasing all the fingerprints of return to a
high-flux, sub-type 1.0 activity state, last observed in mid-1990s.
Importantly, we note a drastic and sudden increase in the flux of  the SXE within a roughly half-year period, possibly driven by
structural changes in the disk/warm corona.  Meanwhile, the increase
in ionizing luminosity has induced emission line responses from a range of
ionization zones in the BLR.
As the community continues to monitor Mkn~590's near-future activity,
including the emission responses from its various circumnuclear
structures, we hope that our results -- e.g., SXE strength, emission
line fluxes as of Fall 2024 -- will serve as a reference guide for comparison to future observations.



\section*{Acknowledgments} 
We would like to thank the anonymous referee for their comments that improved the quality of this manuscript. 
We thank Benny Trakhtenbrot for providing the optical spectrum.
This research was partially supported by Narodowe Centrum Nauki (NCN) grants 2021/41/B/ST9/04110 and 2018/31/G/ST9/03224. M.\'S acknowledges the support from the European Research Council (ERC) under the European Union's Horizon 2020 research and innovation program (grant agreement number 950533), and the Israel Science Foundation (grant number 1849/19).
This work makes use of observations from the Las Cumbres Observatory network. The LCO team is supported by NSF grants AST-1911225 and AST-1911151.

\section*{Data Availability}
Data generated in this research will be shared on reasonable request to the corresponding author.

\bibliographystyle{mnras}

\bibliography{finalv3}



\appendix
\section{Modeling of Optical spectrum}
\label{App:A}

\begin{figure}
   \centering
   
    \includegraphics[height=6.3cm,width=17.0cm]{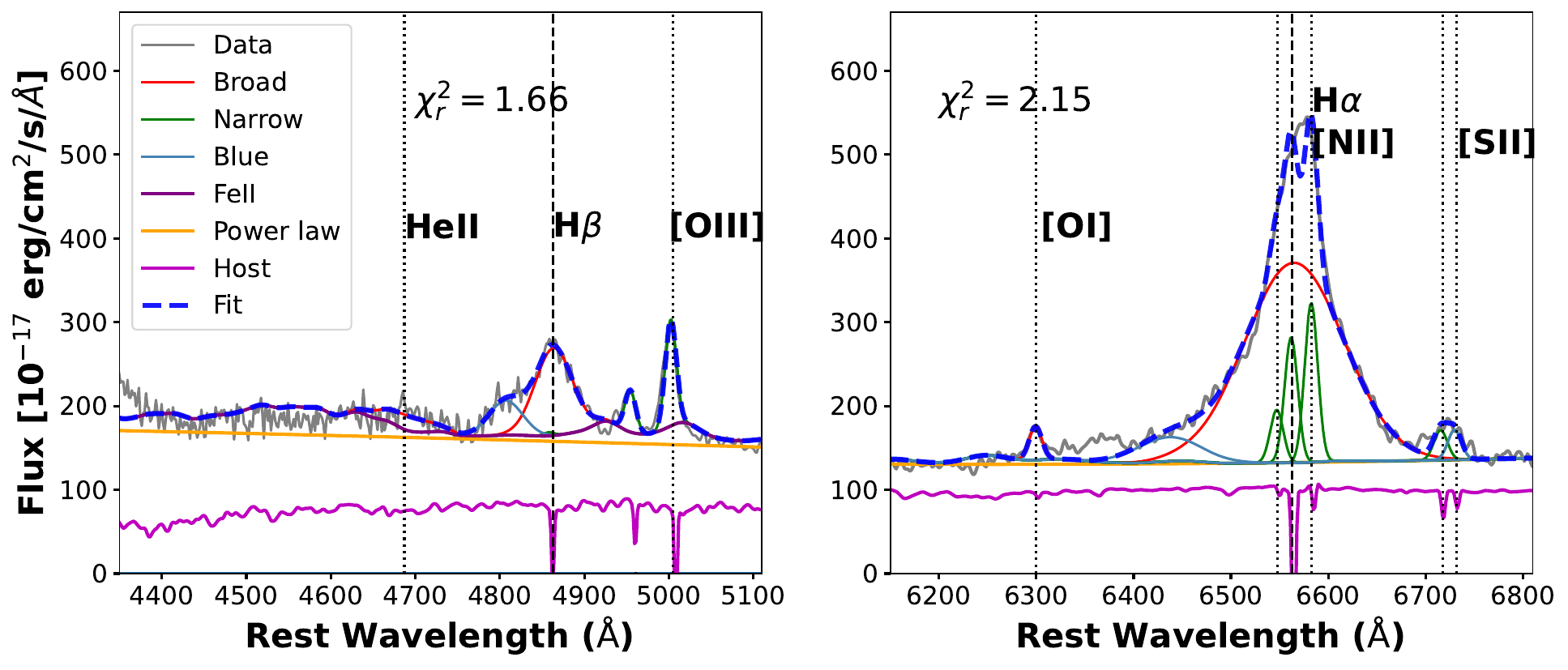}

\caption{Decomposition of LCO Faulkes spectrum using {\sc PyQSOFit} for \hb\ (left panel) and \ha\ (right panel). Details of spectral fitting are given in Sect.~\ref{sec:lco-floyds}.}
    \label{fig:SF}
\end{figure}

\begin{table}
\centering
{\scriptsize
\caption{Best-fitting parameters for fits to the \hb\ and \ha\ regions of both optical spectra. Top: Columns are as follows: instrument, MJD, FWHM of lines in km s$^{-1}$, flux of the Gaussian in $10^{-15}\,{\rm erg\,cm}^{-2}$, $\log L_{5100}$ in erg s$^{-1} $. Bottom:  instrument, MJD, FWHM of lines' blueshifted components in km s$^{-1}$, centroid at the half intensity in km s$^{-1}$.}
\label{tab:opt-spec-measurements}

\begin{tabular}{lllllllll}
\hline
\hline
Name & MJD &  FWHM(\ha\ BC) &  F(\ha\ BC)   &  FWHM(\hb\ BC) &  F(\hb\ BC)    & FWHM(He II) & F(He II) &$\log L_{5100}$ \\ \hline
SDSS	&	52649	&	3900 $\pm$ 100				& 167 $\pm$ 3			&	3500 $\pm$ 200		&	45 $\pm$ 1	&	4000 $\pm$500	&	7$\pm$ 1		& 43.11	\\
LCO	&	60622	&	5500 $\pm$ 100				& 1521 $\pm$ 1			&	3200 $\pm$ 100		&	287 $\pm$ 3	&	4700 $\pm$ 100	&	69	$\pm$ 7	& 43.78  \\ \hline	
Name	&	MJD	&	FWHM(\ha\ BL) &	F(\ha\	BL)	&	c($\frac{1}{2}$)\ha\ BL	&	FWHM(\hb\	BL)	& F(\hb\ BL) & c($\frac{1}{2}$)\hb\	BL		&					\\	\hline	
SDSS	&	52649	&	4200 $\pm$ 100				& 100 $\pm$ 1			&	$-$ 3800 $\pm$ 600			& 3300 $\pm$ 200		& 28 $\pm$ 1	& $-$ 3700 $\pm$ 500		&		\\
LCO	&	60622	&	3700 $\pm$ 100				& 120 $\pm$ 2			&	$-$ 5900 $\pm$ 400			& 2900 $\pm$  100		& 103 $\pm$ 1	& $-$ 3400 $\pm$ 300	\\	\hline		
\end{tabular}
}
\end{table}
\label{lastpage}	
\end{document}